# Empty Substrate Integrated Waveguide-Fed MMW Aperture-Coupled Patch Antenna for 5G Applications

Zia Ullah Khan[1]*, Syeda Fizzah Jilani[1], Angel Belenguer[2], Tian Hong Loh[3], Akram Alomainy[1]

[1] School of Electronic Engineering and Computer Science, Queen Mary University of London, London E1 4NS, UK
[2] Departamento de Ingeniería Eléctrica, Electrónica, Automática y Comunicaciones, Universidad de Castilla-La Mancha, Spain
[3] Engineering, Materials & Electrical Science Department, National Physical Laboratory, Teddington TW11 0LW, UK

*zia.khan@qmul.ac.uk*

*Abstract*—This paper introduces a microstrip square patch antenna with a novel low-cost and efficient feeding mechanism for millimetre-wave frequencies targeting at 28 GHz. The radiating patch is excited by the technique of empty substrate integrated waveguide, and two different slot configurations have been examined in this regard. It has been observed that the antenna gain profile and radiation efficiency is significantly improved in both configurations as compared to the microstrip square patch antenna provided with conventional feeding.

## I. Introduction

The progress of wireless communications and the rapid growing demands of bandwidth have encouraged the researchers to look beyond lower frequencies of currently deployed wireless spectrum. Millimetre-waves (MMW) are considered as the most promising candidate for next generation mobile networks (i.e. 5G) in order to deliver high data throughput and better coverage [1, 2]. Antenna design is a versatile and diverse domain comprising of a number of antenna topologies and configurations. Among these, microstrip patch antennas are preferred due to the planar geometry, simplicity of design and ease of integration. It is observed that the conventional feeding networks introduced for patch antennas suffer mainly due to increased dielectric and radiation losses especially at MMW frequencies. Metallic rectangular waveguides (RWG) are extensively investigated as a feeding network for patch antennas [3]; yet limited due to increased weight and fabrication cost of the structure. In order to deal with such issues Substrate Integrated Waveguide (SIW) structures are developed [4].

A SIW structure is composed of two rows of metalised vias, embedded in dielectric substrate with the conductor cladding on the top and bottom surface of the dielectric [5]. It can also be considered as a dielectric-filled rectangular waveguide, provided with a compact, high quality factor (Q), and low-loss structure [6]. However, the main limitation in this scenario is the presence of dielectric which increases the losses as compare to conventional RWG, especially at higher frequencies. The desired outcome would be to sustain the advantages of RWG without compromising the low profile characteristics and low-cost fabrication suggested by SIW structures. In order to achieve this objective, a simple and low-cost structure is proposed in [7], referred as Empty Substrate Integrated Waveguide (ESIW). ESIW structure is constructed by creating a void or aperture of rectangular shape in the geometry of a planar substrate. Moreover, the upper and lower surfaces as well as the lateral walls are metallised to create an empty-waveguide-like geometry. The results suggest much lower losses and improved *Q*-factor could be achieved in the devices integrated with this novel technique as compared to the equivalent designs in SIW [7]. In this paper, ESIW structure is integrated as a low-loss and low-cost feeding mechanism in a patch antenna for the first time. The purpose of this research is to improve the radiation efficiency as compared to traditional planar feeding structure as well as SIW feeding. Both slot configurations, i.e., transverse and longitudinal, are investigated and results are provided based on numerical evaluation.

## II. Proposed Antenna Design

The designed geometry of ESIW-fed aperture-couple microstrip patch antenna at designed frequency of 28 GHz is presented in Fig. 1. The structure comprises mainly of two substrates assembled in a stacked configuration. The lower substrate is incorporated with ESIW structure, which is utilised to act as a feeding network for a square patch antenna. The patch antenna is mounted on the top surface of second substrate. The patch is excited by a slot etched on the top surface of ESIW. Figures 1(a) and 1(b) depict two slot configurations i.e. transverse or longitudinal, implemented in the antenna design for investigation. The ESIW structure is employed on Rogers 4003 substrate ($h_{ESIW}$ = 0.508 mm, permittivity ($\varepsilon_r$) = 3.55, and cladding thickness t = 35 μm). The design width of ESIW is selected based on a standard $K_a$-band rectangular waveguide (WR-28).

The design procedure for the proposed feed structure is thoroughly demonstrated in [8]. Same parameters for $K_a$-band referred in [8] are implemented in this work. The ESIW structure has the microstrip-to-empty-waveguide transition at one end, and terminated by short-circuit at the opposite end to keep the standing waves inside, shown in Fig. 1 (c). The designed slot is positioned at the peak of the standing waves profile in order to excite the patch at maximum coupling. The slot is also aligned with the centre of the patch. Rogers RT/Duroid 5880 is used for the patch substrate ($h_{PATCH}$ = 0.508 mm, $\varepsilon_r$ = 2.2, t = 35 μm). The optimised dimensions of design parameters of the proposed antenna geometry referred in Fig. 1 are illustrated in Table 1.

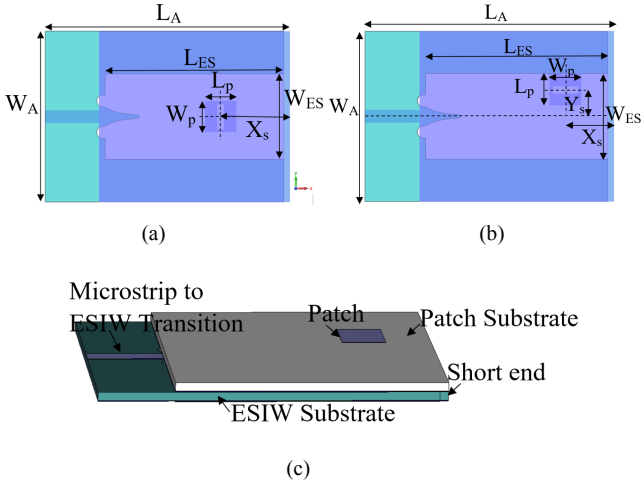

(c)

Fig. 1. Proposed ESIW based aperture-coupled patch antenna: (a) with transverse slot (b) with longitudinal slot (c) side view.

TABLE I. DIMENSIONS OF PROPOSED ESIW BASED PATCH ANTENNA

| Modes | Parameters | mm |
|---|---|---|
| Transverse Slot | Length of Antenna Profile, $L_A$ | 23.5 |
| | Width of Antenna Profile, $W_A$ | 14.2 |
| | Patch length/width, $L_P = W_P$ | 2.4 |
| | Slot Length, $S_L$ | 2.2 |
| | Slot Width, $S_W$ | 1 |
| | Length of ESIW, $L_{ES}$ | 15 |
| | Position from Short end, $X_S$ | 7.6 |
| | Position from Centre axis, $Y_S$ | 0 |
| Longitudinal Slot | Length of Antenna Profile, $L_A$ | 18.5 |
| | Width, $W_A$ | 14.2 |
| | Patch length/width, $L_P = W_P$ | 2.4 |
| | Slot Length, $S_L$ | 2.2 |
| | Slot Width, $S_W$ | 1 |
| | Length of ESIW, $L_{ES}$ | 11.2 |
| | Position from Short end, $X_S$ | 3.8 |
| | Position from Centre axis, $Y_S$ | 2.2 |

## III. RESULTS AND DISCUSSION

The proposed antenna is designed and numerically evaluated using CST Microwave Studio software to inspect the suggested concept of introducing ESIW structure in order to improve the feeding performance of the antenna at 28 GHz. The performance of antenna is inspected based on the S-parameters, radiation pattern, realised gain and efficiency as well as electric field distribution inside the ESIW structure. The obtained results based on simulations are discussed in detail in this section.

### A. S-Paramters

The $S_{11}$ plots of both transverse and longitudinal configurations of designed antenna are provided in Fig. 2. The results show that the designed antenna offers an impedance bandwidth of 6.8% (i.e. 27.4-29.3 GHz) for the transverse slot and 5% (i.e. 27.3-28.7 GHz) for the longitudinal slot configuration.

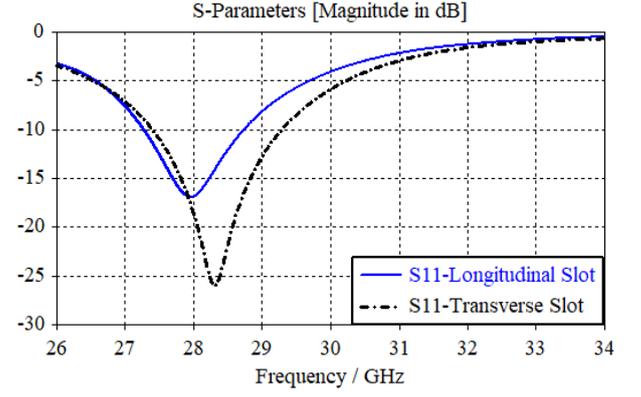

Fig. 2. $S_{11}$ of the proposed ESIW based aperture-coupled patch antenna.

### B. Radiation Pattern

The simulated E- and H-plane radiation patterns computed at 28 GHz are presented in Fig. 3. It is observed that the radiation profile of the antenna is directed along the broadside direction in both of the configurations.

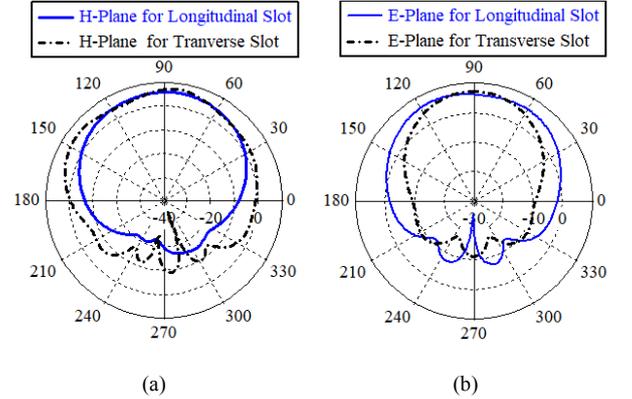

Fig. 3. Radiation pattern of the proposed ESIW based aperture-coupled patch antenna at 28 GHz: (a) H-Plane at Phi=0º, (b) E-Plane at Phi=90º.

### C. Realised Gain and Efficiency

The realised gain and efficiency plots vs. frequency are provided in Fig. 4. The results depict that the realised gain for the square patch fed by the transverse slot is 7.38 dBi and that of a longitudinal slot is 6.57 dBi at 28 GHz. Additionally, numerically computed radiation efficiency of the designed antenna at 28 GHz is approx. 95% due to incorporation of ESIW structures.

### D. Electric Field Distribution

Fig. 5 (a) and (b) present the electric field distribution of the ESIW feeding network and of the radiation patch excited by transverse and longitudinal slot configuration, respectively. It can be observed that the field accelerates forward along the ESIW feeding network and the electromagnetic power is delivered to the patch by the slot.

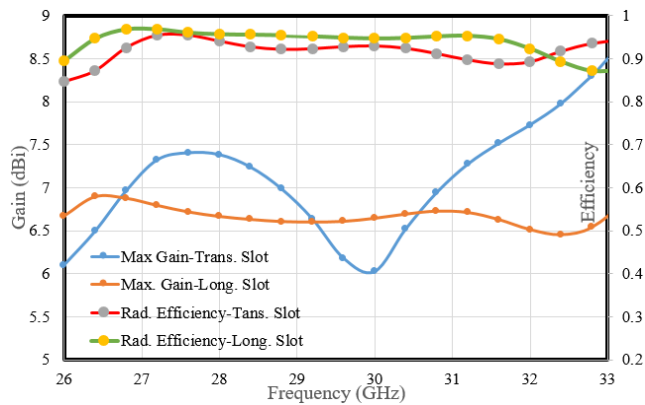

Fig. 4. Realised gain and efficiency vs. frequency of the proposed ESIW based aperture-coupled patch antenna.

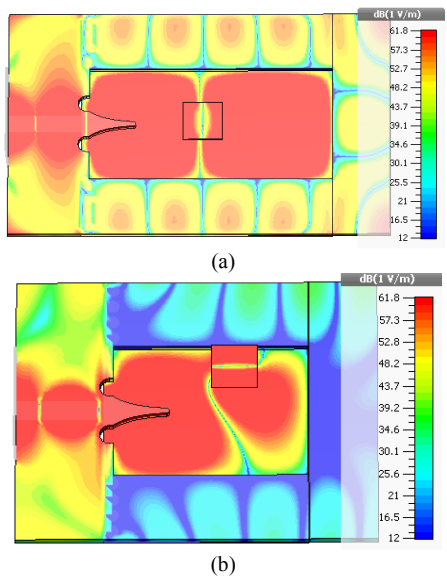

(a)

(b)

Fig. 5. Electric field distribution of the proposed ESIW based aperture-coupled patch antenna.

## IV. CONLUSION

In this paper, a novel technique of ESIW has been introduced as a feeding structure for aperture-coupled patch antenna at 28 GHz. The design is equipped with a low-loss and highly efficient feeding scheme which enhances the antenna performance in a cost-effective way with less complexity involved in the planar design and integration. The designed antenna suggests an operating bandwidth of 6.8% (i.e. 27.4-29.3 GHz) in the transverse slot mode while 5% (i.e. 27.3-28.7 GHz) in the longitudinal slot scenario. High gain of 7.38 dBi and 6.57 dBi are observed in transverse and longitudinal modes respectively, while high efficiency is also accomplished in both of the cases. The performance of antenna recommends its potential to be incorporated in future 5G applications.

## ACKNOWLEDGMENT

The work of T. H. Loh was supported in part by the 2017–2018 Electromagnetic Measurement and Technology Programme of the National Measurement Office (an Executive Agency of the U.K. Department for Business, Energy and Industrial Strategy) and in part by the EMPIR project – MET5G (Metrology for 5G communications), under Projects 120288 and 118927, respectively. The MET5G project has received funding from the EMPIR programme co-financed by the Participating States and from the European Union's Horizon 2020 research and innovation programme.

A part of this work is supported by NPRP through the Qatar National Research Fund; Grant 7-125-2-061.